\documentclass[aps,prl,twocolumn,floatfix]{revtex4}

\usepackage{graphicx}
\usepackage{dcolumn}
\usepackage{bm}

\begin{document}


\title{Comment on J.~Fern\'andez et al, 
   ``Requirements and sensitivity analysis for temporally- and spatially-resolved thermometry using neutron resonance spectroscopy,''
   Rev. Sci. Instrum. {\bf 90}, 094901 (2019).}

\date{October 7, 2019 -- LLNL-JRNL-795286} 

\author{Damian C. Swift}
\email{dswift@llnl.gov}
\affiliation{%
   Lawrence Livermore National Laboratory,
   7000 East Avenue, Livermore, California 94551, USA
}
\author{James M. McNaney}
\affiliation{%
   Lawrence Livermore National Laboratory,
   7000 East Avenue, Livermore, California 94551, USA
}

\maketitle

Neutron resonance spectrometry (NRS) offers a unique way to measure the
bulk temperature inside an optically-opaque sample such as a metal,
with temporal resolution suitable to probe inertially-confined states 
of elevated pressure produced by dynamic loading,
but requires an intense neutron pulse and has previously been demonstrated
only with a nuclear spallation source of $o(10^{14})$ neutrons 
and shocked states 
induced by a centimeter-scale projectile (launched by high explosives) 
persisting for the order of a microsecond \cite{Yuan2005,Swift2008}.
We have previously undertaken research to investigate 
laser-generated neutron pulses \cite{Higginson2010}, assessing whether
it would be feasible to use such pulses for single- or few-pulse 
NRS measurements. 
We also considered whether such measurements could be optimized for use
with laser-induced dynamic loading, which would make the synchronization
of the load and probe much more convenient and potentially less expensive 
than with the projectiles and particle accelerator used previously.
Our conclusion,
which we did not publish but did discuss with interested parties,
was that such measurements were not feasible with the $o(10^{10})$ neutrons
demonstrated in our experiments, even if the measurement could be applied
to a resonance of significantly higher energy than the $\sim$20\,eV
resonances used previously \cite{Yuan2005}.

The article by Fern\'andez et al argues that, because of advances made in
laser-produced neutron pulses, it should now be possible to
obtain a temperature measurement by neutron resonance spectrometry
from a single sub-picosecond laser pulse in the 300\,J range
producing $o(10^{11})$ neutrons.
As a point design, the article takes the moderator and detector configuration
used in the previous experiments at the LANSCE accelerator,
and argues that, because a laser-driven neutron source would not produce 
the $\gamma$-ray flash that accompanies the accelerator-driven spallation
event and which can blind the neutron time-of-flight (NTOF) detector,
the detector can be moved much closer to the moderator:
3.1\,m instead of 23\,m.
From standpoint of designing new experimental facilities, 
this would make it possible to make NRS measurements
of dynamically-loaded states with a laser source of neutrons
costing $\sim$\$10M rather than a particle accelerator source costing 
$\sim$\$1B.
The article implies that a future facility would induce transient 
high-pressure states using a second laser, with a pulse $\sim$200\,ns.

The article gives an estimate of the number of neutrons $N$
needed to be detected in the region
of the resonance for a temperature measurement of a given accuracy,
$2/\pi^{1/4}\sqrt N$.
In our own analysis, we deduced that
$N$ should strictly be the number of neutrons that would have been detected
had it not been for the effect of the resonance, i.e. the number
scattered by the resonance multiplied by the efficiency of the detector, 
$\epsilon_d$.
Apart from this subtlety, we arrived at much the same result
that scientifically useful dynamic temperature measurements 
at the percent level require $N\sim 10^3/\epsilon_d$ neutrons.
Again, we did not publish this result (though we did communicate it to
the authors).
The challenge in NRS measurements is that, although pulsed neutron sources are
demonstrably capable of producing several orders of magnitude more neutrons,
it is difficult to moderate enough of them from their source energy to the energy range
of the resonance and pass them through the sample while it is maintained
at an inertially-confined state of elevated pressure, to achieve this number.

We note that short-pulse laser interactions used to accelerate charged particles,
and thereby neutrons, also produce significant levels of electrical noise and
energetic photons. It is not clear whether NTOFs could be moved as
close as 3\,m from the moderator, though they could likely be closer than 23\,m.
In addition, there seems to be no particular reason why the dynamically-loaded
sample needs to be at $\sim$1\,m as in the LANSCE experiments: replacing
particle accelerators, chemical explosives, and projectiles with lasers 
reduces the need to protect the source from blast and fragments,
and so this separation could be reduced, potentially reducing the amount
of time at which the sample would have to be held at pressure while
neutrons across the energy range of the resonance pass through.

More seriously, the design study does not really indicate a path toward
a practical experiment,
and the neutron budget in particular is misleading as presented.
The calculation of increased neutron efficiency by reducing the distance
to the detector needs to consider 
what this means for the size of the dynamically-loaded region.
In order for all neutrons from the moderator emitted toward the detector
to also pass through the sample, as shown in the article (Fig.~1 therein),
the area of loaded sample presented to the neutrons, $A_s$ must lie between
$A_m$ and $A_d$, the facing areas of the moderator and detector respectively.
If these have the same shape, characterized by a scale length
$l\propto\sqrt A$, then the scale length of the loaded sample
\begin{equation}
l_s=l_m+(l_d-l_m)\frac L{L+L_d},
\end{equation}
where $L$ and $L_d$ are the moderator to sample and sample to detector 
separations respectively.
For the example configuration, with $A_m$=56.25\,cm$^2$,
$A_d=$1393\,cm$^2$, $L=$1\,m, and $L_d=$3.1\,m, $A_s\simeq$212\,cm$^2$.
In practice, as shown in the figure in the article, the sample would likely be
oriented with its normal at an angle $\theta$ say to the neutron beam,
in which case the actual loaded area would be at least a factor $1/\cos\theta$
larger.
Putting it mildly,
this would be an impractical area over which to induce dynamic loading
with a laser.
Even the limiting case where the sample is butted up flat against the
moderator would require at least 56.25\,cm$^2$ to be uniformly driven,
which is far in excess of any laser experiment we are aware of.
The cost of a laser capable of such an experiment depends on other
aspects, such as the peak pressure of interest, and whether
it could be achieved by tamped ablation \cite{Fabbro1990,Colvin2003,Paisley2008} 
or whether ablation into vacuum would have to be employed,
but it could certainly exceed the cost of the
National Ignition Facility (NIF) by a factor of several,
and could thus easily exceed \$10B.
Tamped ablation on timescales $o(0.1)\,\mu$s has not,
to our knowledge, been demonstrated for pressures above 20\,GPa
\cite{AkinSwift};
free ablation at 100\,GPa requires $\sim$20\,PW/m$^2$
(or J/ns.mm$^2$), i.e. $\sim$22.5\,MJ for 200\,ns over 56.25\,cm$^2$,
compared with $\sim$2\,MJ available at NIF.
Less costly loading methods might be employed, whether chemical explosives,
projectiles, or electrical pulsed power, but any technique could potentially
risk damage to the short-pulse infrastructure or detectors, 
unless moved further away.
The example design thus does not represent an almost-viable concept,
and any plausible variant would require even higher drive energy or more
neutrons.

The size of the moderator appears to be a fundamental limitation in attempting
to scale down NRS experiments.
Demonstrated techniques for generating neutrons with lasers
(or almost any other technique) involve nuclear reactions with typical
energies in the megavolt range or higher.
The most efficient moderator is the hydrogen nucleus;
hydrogen-rich substances such as polyethylene are cheap choices and there is
relatively little to be gained by looking for more exotic substances with a
slightly higher density of hydrogen.
The mean free path of $O(\mbox{100\,keV-MeV})$ neutrons in polyethylene
is $\sim$0.1\,m.
To reduce the neutron energy from $\sim$1\,MeV to $\sim$20\,eV requires
$\sim$15 scatters, or a random walk of $\sim$4 mean free paths.
Thus, moderating a high proportion of the neutrons requires a moderator
$o(0.1)$\,m in size, which will emit neutrons over an area $o(0.01)$\,m$^2$
as at LANSCE.

The design strategy used at LANSCE was to reduce the effective size of the
neutron source by collimation, sacrificing neutrons.
Another strategy is to use a moderator thinner than the mean free path
to the first scattering event, for instance selecting
neutrons scattered nearly straight back toward the source,
which would have the lowest energy.
This strategy again throws away most of the neutrons, 
but gives both a relatively small size
and a shorter spread in time over the width of any given resonance.
It could possibly be practical with neutron yields $o(10^{16})$ or more
as have been demonstrated at NIF \cite{LePape2018},
but may still need to be combined with a drive pulse of longer duration
than available there.

An avenue for further refinement is to consider higher energy resonances, 
requiring 
less down-scattering of the neutrons and thus a thinner moderator and
shorter minimum duration of the load.
However, these advantages must be offset against potentially less efficient
neutron detectors at higher energies, and larger distances to the 
detector to achieve a given NTOF spectral precision.

In conclusion, the increase in neutron yields from short-pulse laser sources,
demonstrated since our previous study, are not yet adequate for NRS temperature
measurements of dynamically-loaded samples.
Simply bringing the detector closer to the moderator would not help as dynamic
loading then becomes impractical.
Orders of magnitude advances in neutron production are needed before 
short-pulse laser facilities could usefully perform single-shot NRS thermometry.
Nuclear spallation at the LANSCE accelerator, and possibly inertial confinement
fusion at NIF, are as yet the only neutron sources bright enough to
contemplate employing for single-shot thermometry.
LANSCE remains the only facility in which such experiments are
currently feasible, using $\sim$microsecond scale dynamic loading.
It is surprising that greater advantage has not been taken of this
unique capability.

This work was performed under the auspices of
the U.S. Department of Energy under contract DE-AC52-07NA27344.

\end{document}